\documentclass[aps,prd,preprint,showpacs,floats]{revtex4}
\usepackage[dvips]{graphicx}
\usepackage{bm}
\begin{document}
\title{Power Corrections in Charmless B Decays }
 \author{T. N. Pham}
 \email[E-mail address: ]{Tri-Nang.Pham@cpht.polytechnique.fr}
  \author{Guohuai Zhu}
  \email[E-mail address: ]{Guohuai.Zhu@cpht.polytechnique.fr}
 \affiliation{Centre de Physique Theorique, \\
Centre National de la Recherche Scientifique, UMR 7644,  \\  
Ecole Polytechnique, 91128 Palaiseau Cedex, France} 
  \date{\today}
\begin{abstract}
In this paper, we focus on the role of power corrections in QCD
factorization(QCDF) method in charmless two-body nonleptonic $B$ meson decays.
We use the ratio of the branching fraction of $B^+ \to \pi^+ K^{\ast 0}$ to
that of $B^0 \to \pi^- \rho^+$, for which the theoretical uncertainties are
greatly reduced, to show clearly that the power corrections in charmless B
decays are probably large. With other similar ratios considered, for 
example, for the $B^0 \to K^- \rho^+$ decay, it is very 
likely that, among various sources of power corrections, annihilation
topology  plays an indispensable role at least for penguin dominated 
$\rm PV$ channels. We also consider some selective 
ratios of direct CP asymmetries.
Among these, we find that, if power corrections other than the 
chirally enhanced power corrections and annihilation topology 
were negligible, QCDF would
predict the direct CP asymmetry of $B \to \pi^+ \pi^-$ to be 
about 3 times larger than that of
$B \to \pi^\pm K^\mp$, with opposite sign. Experimentally any significant
deviation from this prediction
would suggest either new physics or possibly  the importance of long-distance
rescattering effects.

\end{abstract}
\maketitle
\section{Introduction}
With the excellent performance of KEK and SLAC B factories, a great number
of charmless $B$ decay channels have been measured with high precision. It
is even more exciting that, with the expected larger data sample in the
near future,  CP violations in many decay channels might be soon
reachable. However in most cases, our ignorance of strong dynamics stands
in the way of
either identifying the signal of new physics or extracting the
fundamental Cabibbo-Kobayashi-Maskawa (CKM) angles from the observables. 

Theorists have tried hard to have a better understanding on the strong
dynamics in non-leptonic $B$ decays, in which QCD factorization method
(QCDF)\cite{QCDF} is a recent progress. 
In the heavy quark limit, QCDF has shown
that the hadronic matrix elements of $B$ decays have simpler
non-perturbative structures, while the QCD corrections are perturbatively
calculable, at least at one-loop order. However the $1/m_b$ power
corrections, especially annihilation contributions and chirally enhanced
power corrections \cite{QCDF1,DYZ}, are phenomenologically important. It
is even worse that
factorization generally breaks down beyond the leading power expansion. As
a result, the model-dependent parametrization has to be introduced in
QCDF to account for these formally power-suppressed contributions. For
some decay channels, these power corrections could even
compete with the
leading power contributions in some parameter space which makes the
reliability of QCDF predictions in doubt. Very recently, a new effective
theory, called soft-collinear effective theory (SCET), has been applied to
charmless $B$ decays \cite{SCET}, in which the power corrections can be
studied in a systematic way, though more non-perturbative operators have
to be introduced. However since this method is still under development and
several issues remain to be resolved, we shall not discuss SCET further in
this paper. 

Thus it is of great interest to investigate the role of power corrections in
charmless $B$ decays, and to determine if the power corrections were
really important,
whether they were  mainly from   chirally enhanced corrections
and annihilation topology or  from other sources of power corrections,
such as long-distance rescatterings. 
In this paper, we
shall discuss these problems in a transparent way based on experimental
measurements.

\section{QCD factorization for charmless $B$ decays}
To make the paper self-contained, we will recapitulate the main point
of QCDF in the following. One may find more details in Refs 
\cite{QCDF,QCDF1,DYZ,QCDF2,DYZPV}.

There are three distinct scales: $M_W \gg m_b \gg \Lambda_{QCD}$ involved
in charmless $B$ decays. To go beyond naive model estimation, the physics of
different scales should be separated from each other. It is known that,
with the operator product expansion and renormalization group equation,
the effective Lagrangian can be obtained, in which short-distance effects
involving large virtual momenta of the loop corrections from the scale
$M_W$ down to $\mu ={\cal O}(m_b)$ are cleanly integrated into the Wilson
coefficients. Then the amplitude for the decay $B \to M_1 M_2$ can be
expressed as \cite{Buras}:
\begin{equation}
{\cal A}(B \to M_1 M_2)=\frac{G_F}{\sqrt{2}} \sum_{i=1}^6 \sum_{q=u,c}
\lambda_q C_i (\mu) \langle M_1 M_2 \vert Q_i (\mu) \vert B \rangle ~,
\end{equation}
where $\lambda_q$ is a CKM factor, $C_i (\mu)$ is the Wilson coefficient
which is perturbatively calculable from first principles. The effective
operators $Q_{1,2}$ and $Q_{3,...,6}$ are tree level  and QCD penguin
operators, respectively. We have neglected electroweak penguin operators
$Q_{7,...,10}$ because their effects are numerically small in most decay
channels.  $\langle M_1 M_2 \vert Q_i (\mu) \vert B \rangle$ is a hadronic
matrix element which contains the physics effects from the scale $\mu =
{\cal O}(m_b)$ down to $\Lambda_{\rm QCD}$. Since the perturbative and
nonperturbative effects related to $m_b$ and $\Lambda_{\rm QCD}$ still
entangle with each other, it is highly nontrivial to estimate the hadronic
matrix elements. But in the heavy quark limit, QCDF shows that the above
hadronic matrix elements can be factorized into hard radiative corrections
and simpler nonperturbative structures which can be parametrized by the
form factors and meson light-cone distribution amplitudes (LCDAs).  

But for phenomenological applications, the power corrections in $1/m_b$,
especially the chirally enhanced corrections and annihilation
contributions should be taken into account. Chirally enhanced corrections
arise from two-body twist-3 LCDAs of the  final state mesons. Unfortunately the
factorization breaks down for hard spectator scattering diagrams because
the twist-3 LCDA does not approach zero in the endpoint region which
introduces a logarithmic divergence. A similar divergence also appears in
the annihilation contributions. Phenomenologically, Beneke {\it et al.}
\cite{QCDF1} adopted a model parametrization for the endpoint
divergence:
\begin{equation}
X_{A,H}=\int \limits_0^1 \frac{dx}{x} = \log \frac{m_B}{\Lambda_h} 
(1+\rho_{A,H} e^{i \phi_{A,H} } ) ~~~~~(0 \le \rho_{A,H} \le 1 ) ~,
\end{equation}  
where $X_A$ ($X_H$) denotes the divergent terms from annihilation 
topology (hard spectator scattering). The corresponding price to pay
is model-dependence uncertainties and 
 also large numerical uncertainties from $X_{A,H}$. 
With the above discussions, the decay amplitudes can be written as
\begin{equation}
{\cal A}(B \to M_1 M_2)=\frac{G_F}{\sqrt{2}} \sum_{p=u,c} v_p \left ( 
\sum_{i=1}^6 a_i^p \langle M_1 M_2 \vert O_i \vert B \rangle_f + 
\sum_{j} f_B f_{M_1}f_{M_2} b_j \right ),
\end{equation}   
where $\langle M_1 M_2 \vert O_i \vert B \rangle_f$ is the factorized
hadronic matrix element which has the same definition as that in the naive
factorization approach. For the complete expressions of QCD
coefficients $a_i$ and annihilation parameters $b_j$, one may refer to 
Ref. \cite{QCDF2}.

\section{Are power corrections large ?}

It is clear that chirally enhanced power corrections should be large, at 
least in $B$ decays to two light pseudoscalar mesons, because although
they   
are suppressed by the factor (taking the pion as an illustration)
\begin{equation}
r_\chi^\pi = \frac{2 m_\pi^2}{m_b (m_u + m_d)} \mbox{~,}
\end{equation}  
which is formally of the order of $\Lambda_{QCD}/m_b$, numerically this
factor
is of order one. Experimentally, the large branching ratios of $B \to \pi K$
decays also strongly support this statement. For example, the ratio
\begin{equation}
\frac{{\cal B}(B^+ \to \pi^+ K^0)}{{\cal B}(B^+ \to \pi^+ K^{\ast 0})}=
\frac{(21.8 \pm 1.4) \times 10^{-6}}{(9.0 \pm 1.3) \times 10^{-6}}=2.4 \pm
0.4
\end{equation} 
would be theoretically smaller than $1$ without chirally enhanced
contributions, say $r_{\chi}^K a_6$ term. The experimental data in the
above equation are from ref. \cite{HFAG1}.  

However it is not very clear about the role of annihilation topology 
in $B$ decays. It was once believed to be quite small in most $B$ decay
channels,  because of the power suppression \cite{Ali}. The importance of
annihilation contributions was first noticed in the perturbative QCD
method for charmless $B$  decays\cite{pQCD}. Recent phenomenological
analyses 
\cite{DYZGA,QCDF2,aleksan,cottingham} based on QCDF also suggest 
substantial
contributions from annihilation topology. However since in QCDF, 
there are many
parameters involved in the global analysis of experimental data, it would
be interesting to show the importance of annihilation topology in a 
transparent way. In the following we will try to do it with various
ratios. We find that annihilation topology very likely plays a significant
role, at least in penguin dominated
$B \to PV$ decays, where $P$ denotes light pseudoscalar meson and $V$
denotes light vector meson. 

Let us first consider the ratio of the decay amplitude of $B^+
\to \pi^+ K^{\ast 0}$ to that of $B^0 \to \rho^+ \pi^-$. If the power
corrections were negligible, this ratio would be theoretically very
clean where the form factors cancel out, furthermore it is
almost independent ot the  CKM angle $\gamma$ and the strange-quark mass:
 \begin{equation} \label{pik}
\left \vert \frac{{\cal A}(B^+ \to \pi^+ K^{\ast 0})}{{\cal A}(B^0 \to
\rho^+ \pi^-)}\right \vert \simeq 
\left \vert \frac{ V_{cb}V_{cs}}{V_{ub}V_{ud}} \right \vert
\frac{f_{K^\ast}}{f_\rho} \left \vert 
\frac{a_4^c (\pi K^\ast)+r_\chi^{K^\ast} a_6^c (\pi K^\ast) }{a_1^u}
\right \vert \mbox{~,}
\end{equation}
where the penguin contributions to $B^0 \to \rho^+ \pi^-$ decay 
and the term proportional to $V_{ub}V_{us}$ in the numerator are
neglected. This should be a reasonable approximation up to a few percent
level. In QCDF, $\vert (a_4^c (\pi K^\ast)+r_\chi^{K^\ast} a_6^c (\pi
K^\ast) ) /a_1^u \vert$ should be about or less than 0.04 (With the 
default parameters in ref. \cite{QCDF2}, it is $0.03$)
and $f_{K^\ast}/f_\rho$ is very close to unity. While for the CKM matrix
elements, there is a useful inequality \cite{Pham,Buras1}: 
\begin{equation}\label{inequality}
\left \vert \frac{V_{ub}}{V_{cb}}\right \vert = \lambda \sin \beta 
\sqrt{1+\frac{\cos^2 \alpha}{\sin^2 \alpha}} \ge \lambda \sin \beta ~.
\end{equation}   
The current measured value of $\sin 2\beta$ from BaBar and Belle gives
\cite{HFAG}
\begin{equation}
\sin 2\beta= 0.736 \pm 0.049 \mbox{~,}
\end{equation}
from which we obtain $\sin \beta = 0.402 \pm 0.033$. In principle we could
get another solution $\sin \beta \simeq 0.9$ which however is inconsistent
with the direct $\vert V_{ub} \vert$ measurements. So given 
$\lambda=0.224$, we will get an interesting lower limit 
\begin{equation}\label{inequality2}
\left \vert \frac{V_{ub}}{V_{cb}} \right \vert \ge \lambda \sin \beta =
0.090 \pm 0.007 > 0.078 ~~~~~ \mbox{( 90\% C.L.) .} 
\end{equation}
With this lower limit, there arises a clear discrepancy
between theory and experiments: 
\begin{equation}\label{pikrhopi}
0.53 > \left \vert \frac{{\cal A}(B^+ \to \pi^+ K^{\ast
0})}{{\cal A}(B^0 \to \rho^+ \pi^-)}\right \vert = \left [ 
\frac{\tau(B^0)}{\tau(B^+)} \frac{{\cal B}(B^+ \to \pi^+ K^{\ast  
0})}{{\cal B}(B^0 \to \rho^+ \pi^-)} \right ]^{1/2} = 0.77 \pm 0.09~,
\end{equation}
where the left hand side is from theoretical estimation Eqs.
(\ref{pik}), (\ref{inequality}), (\ref{inequality2}) and the right
hand side is obtained from the following experimental measurements: 
${\cal B}(B^0 \to \rho^+ \pi^-)=13.9 \pm 2.7$ \cite{laget}, 
${\cal B}(B^+ \to \pi^+ K^{\ast 0})=9.0 \pm 1.3$ \cite{HFAG1} and 
$\tau(B^+)/ \tau(B^0)=1.083 \pm 0.017$ \cite{PDG}.
Notice that $\cos^2 \alpha$ is of the order of a few percent, it seems
appropriate to neglect it in Eq. (\ref{inequality}). Then the left hand
side of Eq. (\ref{pikrhopi}) would be reduced further to $0.46 \pm 0.04$
if we fix $(a_4+r_\chi a_6)/a_1$ to be $0.04$. In fact, $(a_4+r_\chi
a_6)/a_1$ for
$\pi K^\ast$ channels is typically less than 0.04 in most of the QCDF
parameter space. For example, if using the default parameters in ref.
\cite{QCDF2}, we will get $0.35$ for this ratio.
Since the chirally enhanced corrections
for this ratio are not expected to be large because 
$a_6 (\pi^+ K^\ast)$ vanishes at tree level, this is a strong indication
that annihilation topology and/or other sources of power corrections might 
play an important role at least in $B \to PV$ decays.  

If we turn to another ratio, the branching fraction of $B^0 \to K^+
\rho^-$ to that of $B^0 \to \rho^- \pi^+$, there is similar disagreement,
although this time it is not well established considering
the large theoretical uncertainties. Assuming
negligible annihilation contributions, the form factors cancel out again
for this ratio. However there are significant dependences on angle
$\gamma$, $V_{ub}/V_{cb}$ and especially the strange-quark mass $m_s$ in
the numerator because of the factor $r_\chi^K$:
\begin{equation}
\left \vert \frac{{\cal A}(B^0 \to K^+ \rho^-)}{{\cal A}(B^0 \to
\rho^- \pi^+)}\right \vert \simeq 
\frac{f_K}{f_\pi} \left \vert \lambda e^{i \gamma}+\left \vert 
\frac{V_{cb}}{V_{ub}} \right \vert 
\frac{a_4^c (\rho K)-r_{\chi}^K a_6^c (\rho K) }{a_1^u}
\right \vert \mbox{~,}
\end{equation} 
where as usual, the small penguin terms in the denominator has been
neglected. If we take the default parameters in ref. \cite{QCDF2}, namely
$\gamma=70^\circ$, $V_{ub}/V_{cb}=0.09$ and $a_4^c (\rho K)-r_\chi^K a_6^c
(\rho K)=0.037+0.003 i$
with the strange-quark mass set to be $90$ MeV, the theoretical
prediction for the ratio 
${\cal B}(B^0 \to K^+ \rho^-)/{\cal B}(B^0 \to \rho^- \pi^+)$ is
$0.38$, which is again significantly smaller than the experimental
measurements $1.01 \pm 0.34$ \cite{HFAG1,laget}. Of course, when the
theoretical uncertainties, especially the uncertainty of the strange-quark
mass, are taken into account, the disagreement is not that impressive. For
example, the ratio can rise up to 0.69 when $m_s$ is lowered to $70$ MeV.  

However when combining  with the $B^+ \to \pi^+ K^{\ast 0}$ decay, this 
result shows that there
is a tendency that, the penguin-dominant $B \to PV$
decay amplitudes are consistently underestimated without annihilation 
contributions. When they are included, 
$B^+ \to \pi^+ K^{\ast 0}$ is dominated by $f_{K^\ast}F^{B \pi}m_B^2 a_4
+b_3(V, P)$ while $B^0 \to K^+ \rho^-$ is dominated by 
$f_K A^{B\rho}_0 m_B^2 (a_4-r_\chi^K a_6)+b_3(P,V)$ with 
\begin{equation}
b_3 (M_1, M_2)=\frac{C_F}{N_c^2}\{ C_3 A_1^i (M_1, M_2) + C_5 
  A^i_3 (M_1, M_2) + (C_5+N_c C_6) A^f_3 (M_1, M_2) \}~.
\end{equation}
Since the penguin
terms $a_4 \simeq -0.03$ and $a_4-r_\chi^K a_6 \simeq 0.037$ are of 
opposite sign, the key observation is that, the annihilation terms 
$b_3 (V, P)$ and $b_3(P, V)$ are also roughly of the opposite sign,
because numerically $(C_5+N_c C_6) A^f_3$ is the dominant 
term in $b_3$ while $A^f_3(P,V)=-A^f_3(V,P)$ if the annihilation
parameter $X_A$ are the same for both channels. So with the inclusion of
annihilation topology, QCDF can easily enhance both ratios without fine
tuning. For example, using the default parameters given in ref.
\cite{QCDF2} but letting the annihilation parameter $\rho_A =1$ 
we can see from Fig. 1 that the ratios are then consistent with
 experimental observations. We believe that this is a strong indication
that annihilation topology probably plays an important role, at least in
penguin-dominated $B \to PV$ decays.  
\begin{figure}[htb]
\begin{center}
\unitlength 1mm
\begin{picture}(160,78)
\put(0,0){\includegraphics{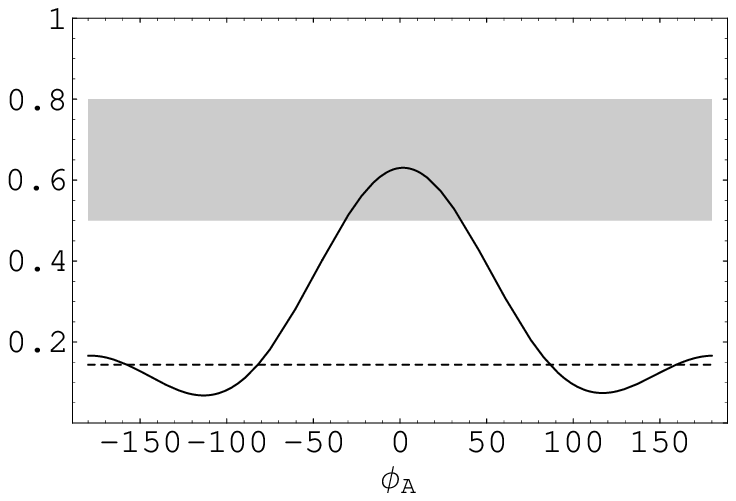}}
\put(80,0){\includegraphics{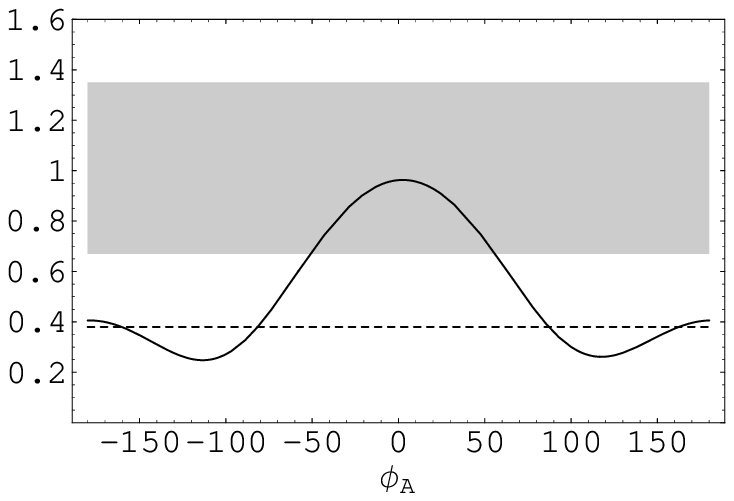}}
\end{picture}
\caption{ The ratios ${\cal B}(B^+ \to \pi^+ K^{\ast 0})/  
 {\cal B}(B^0 \to \rho^+ \pi^-)$ (left plot) and 
${\cal B}(B^0 \to K^+ \rho^-)/{\cal B}(B^0 \to \rho^- \pi^+)$ (right plot)
versus the weak annihilation phase
$\phi_A$. The default parameters of ref. \cite{QCDF2} are used but letting
the annihilation parameter $\rho_A=1$. The dashed lines show the
ratios without weak annihilation contributions. The gray areas denote
the experimental measurements with $1 \sigma$ error. }
\end{center}
\end{figure}  

Similar disagreement can be found in the ratio of 
$B^0 \to \omega K^0$ over $B^0 \to \pi^+ \rho^-$. But we will not go into
details here because $B^0 \to \omega K^0$ decay is quite similar to 
$B^0 \to K^+ \rho^-$ decay. It may also be interesting to discuss 
$B \to \phi K$ decays, but again they are quite similar 
to $B \to \pi K^\ast$ decays if there is no new physics contributions 
in this decay.  

\section{Direct CP violations}
If one agrees that annihilation terms are very likely important in
charmless $B$ decays, one negative point is that, these intractable terms
will bring large uncertainties to the theoretical predictions, not
only in the  branching fractions, but also in  direct CP asymmetry . 
In the following
we will try to see whether we can get some interesting relations with a
few selective ratios of direct CP asymmetry (DCPVs).

Let us first consider the ratio of DCPVs
$A_{CP}(\pi^+ \pi^-)$ over $A_{CP}(\pi^- K^+)$. Belle has claimed large
DCPV observed in $B^0 \to \pi^+ \pi^-$ decay while BaBar has not confirmed
it yet, but both of them are close to a measurement on $A_{CP}(\pi^- K^+)$
\cite{Belle,BaBar}
\begin{equation}
A_{\pi\pi}=\left \{ \begin{array}{ll}
0.58 \pm 0.15 \pm 0.07 & \mbox{(Belle)}~, \\
0.19 \pm 0.19 \pm 0.05 & \mbox{(BaBar)}~;
\end{array} \right. 
A_{\pi K}=\left \{ \begin{array}{ll}
(-8.8 \pm 3.5 \pm 1.8)\% & \mbox{(Belle)}~, \\
(-10.7 \pm 4.1 \pm 1.2)\% & \mbox{(BaBar)}~.
\end{array} \right. 
\end{equation}  
Since QCDF generally predicts small DCPVs for charmless $B$ decays, at
first glance it seems to contradict the Belle data badly. But notice that 
annihilation topology and other sources of power corrections could bring
large uncertainties, it is of interest to have further investigations. In
QCDF, we could define the tree and penguin amplitudes as:
\begin{eqnarray}
T_{\pi \pi}&=&f_\pi F^{B\pi}m_B^2 (a_1^u + a_4^u +r_\chi a_6^u)+
f_B f_\pi f_\pi (b_1+b_3+2 b_4) \equiv f_\pi F^{B\pi}m_B^2 T~, 
\nonumber \\
P_{\pi \pi}&=&f_\pi F^{B\pi}m_B^2 (a_4^c+r_\chi a_6^c)+f_B f_\pi f_\pi 
(b_3+2 b_4) \equiv f_\pi F^{B \pi}m_B^2 P ~,\nonumber \\
T_{\pi K}&=&f_K F^{B\pi}m_B^2 (a_1^u + a_4^u +r_\chi a_6^u)+
f_B f_\pi f_K b_3 \equiv f_K F^{B\pi}m_B^2 {\tilde T}~, \nonumber \\
P_{\pi K}&=&f_K F^{B\pi}m_B^2 (a_4^c+r_\chi a_6^c)+f_B f_\pi f_K
b_3 \equiv f_\pi F^{B \pi}m_B^2 {\tilde P}~. 
\end{eqnarray}
Then the DCPVs can be expressed as
\begin{eqnarray}
A_{\pi \pi} &\equiv& \frac{{\cal B}({\bar B}^0 \to \pi^+ \pi^-)-
 {\cal B}(B^0 \to \pi^+ \pi^-)}{{\bar B}^0 \to \pi^+ \pi^-)+   
 {\cal B}(B^0 \to \pi^+ \pi^-)}=\frac{4 \vert V_{ub}V_{ud}
V_{cb}V_{cd} T P\vert \sin \gamma \sin \delta }
{2 {\cal B}(B \to \pi^+ \pi^- )} ~\mbox{,} \nonumber \\
A_{\pi K} &\equiv& \frac{{\cal B}({\bar B}^0 \to \pi^+ K^-)-   
 {\cal B}(B^0 \to \pi^- K^+)}{{\bar B}^0 \to \pi^+ K^-)+    
 {\cal B}(B^0 \to \pi^- K^+)}=-\frac{4 \vert V_{ub}V_{us} 
V_{cb}V_{cs} {\tilde T} {\tilde P}\vert \sin \gamma \sin {\tilde \delta} }
{2 {\cal B}(B \to \pi^+ K^- )} ~\mbox{,}
\end{eqnarray}  
where $\delta=\delta_P - \delta_T$ is the strong phases difference between
the penguin and tree amplitudes. It is easy to see that many factors
cancel out for the ratio
\begin{equation}\label{DCPV}
\frac{A_{\pi\pi}}{A_{\pi K}}=-\frac{f_\pi^2}{f_K^2}
\frac{{\cal B}(B \to \pi^+ K^- )}
{{\cal B}(B \to \pi^+ \pi^- )} \left \vert \frac{TP}
{{\tilde T} {\tilde P}} \right \vert 
\frac{\sin \delta}{\sin {\tilde \delta}}
\simeq (-2.7 \pm 0.3) 
\frac{\sin \delta}{\sin {\tilde \delta}} ~,
\end{equation}
where the ratio $TP/{\tilde T} {\tilde P}$ has been taken to be $1$, which
is a reasonable approximation in QCDF at about $10$ percent level
uncertainty. The experimental data on relevant branching ratios
\cite{HFAG1} have been used in the above equation, and only the
experimental uncertainties are included in the error estimation. 

At first sign, it is amazing to realize that one would expect very 
naturally a larger DCPV
for $\pi^+ \pi^-$ decay compared with $\pi^- K^+$ decay, if the strong
phases of both channels are not too small and of similar magnitude.
Notice that the DCPV measurements already told us that at least the strong
phase of $\pi^+ \pi^-$ channel should not be small and QCDF predicts quite
similar strong phases for both channels if other sources of power
corrections are negligible.
Actually, it is quite plausible that CP-violating asymmetry in 
$\pi^+ \pi^-$ decay should be bigger than that for $\pi^- K^+$ by a
factor of $3-4$ since the $\pi^+ \pi^-$ decay rate is smaller than
the $\pi^- K^+$ decay rate by a similar factor $3-4$, given the fact
that in QCDF the CP-violating parts of the tree-penguin interference
terms are almost equal as shown above. 
The current experimental value \cite{Belle,BaBar} for this
ratio is 
\begin{equation}
\frac{A_{\pi\pi}}{A_{\pi K}}=\frac{0.42 \pm 0.13}{-0.10 \pm 0.03}
  =-4.2 \pm 1.8 ~,
\end{equation}
which is still consistent, within $1 \sigma$ error, with the theoretical
estimation of $-2.7 \pm 0.3$ under the assumption that the strong phases
$\delta$ and ${\tilde \delta}$ are the same. It should be interesting to
keep an eye on this ratio and any significant deviation from theoretical
estimation would suggest either different strong phases between $\pi \pi$
and $\pi K$ decays or New physics effects affecting one of the channels.
It is also interesting to note that inelastic 
$B \to DD \to \pi \pi$ and $B \to DD_{s} \to K \pi$ rescatterings or
charming penguin contributions could
also produce a large strong phase, but since these contributions are
related by $SU(3)$ symmetry and CKM factor, the interference terms
would be essentially equal and our relation for the CP asymmetry applies
and we expect large CP asymmetry in $B \to \pi\pi$ as found
recently \cite{Isola,Ciuchini}.

With the same reasoning, we can get similar relations for other
decay channels such as:
\begin{eqnarray}
\frac{A_{CP}(B^0 \to \rho^+ \pi^-)}{A_{CP}(B^0 \to K^{\ast +} \pi^-)}
&\simeq& -\frac{{\cal B}(B^0 \to K^{\ast +} \pi^-)}{{\cal B}(B^0 \to
\rho^+ \pi^-)}\frac{f_\rho^2}{f_{K^\ast}^2}\frac{\sin \delta_{\pi \rho}}
{\sin \delta_{\pi K^\ast}}\mbox{~,} \nonumber \\
\frac{A_{CP}(B^0 \to \rho^- \pi^+)}{A_{CP}(B^0 \to \rho^- K^+ )}
&\simeq& -\frac{{\cal B}(B^0 \to \rho^- K^+ )}{{\cal B}(B^0 \to 
\rho^- \pi^+)}\frac{f_\pi^2}{f_{K}^2}\frac{\sin \delta_{\rho \pi}}
{\sin \delta_{\rho K}}\mbox{~.} 
\end{eqnarray}
If the above pairs of strong phases are roughly the same, which is true in
QCDF, The DCPVs of the penguin-dominated decays would be about 1.5 times
larger than those of their tree-dominated counter parts. Precise
measurements on these ratios could help us to get some insight into the 
strong dynamics of $B$ decays. 
  
\section{Conclusion}
In this paper, we have used the ratios of the branching fraction of $B^+
\to \pi^+ K^{\ast 0}$ to that of $B^0 \to \pi^- \rho^+$, and of  
$B^0 \to K^+ \rho^-$ to that of $B^0 \to \pi^- \rho^+$, to show clearly
and with greatly reduced theoretical uncertainties,
that  power corrections in charmless B
decays are probably large. The key observation is that QCDF predicts
the annihilation terms for $B^+\to \pi^+ K^{\ast 0}$ and $B^0 \to K^+ \rho^-$
are almost the same magnitude but opposite in sign. The 
result is that both these two branching fractions could be enhanced by
the annihilation contribution to accommodate the experimental data
without fine tuning. Assuming the presence of the annihilation
contribution, we then derive relations between CP asymmetries for
a few selective decay channels and shows that QCDF would
predict the direct CP asymmetry
of $B \to \pi^+ \pi^-$ to be about 3 times larger than that of
$B \to \pi^\pm K^\mp$ with opposite sign, which is consistent, within one
sigma error, to the
current experimental data $-4.2 \pm 1.8$.  
Any significant deviation from this prediction
 would suggest either new physics or possibly the importance 
of long-distance 
rescattering effects. We also discuss other similar direct CP ratios which
might help us to get some insight into the strong dynamics of charmless B
decays.

\section*{Note Added}
After finishing this paper, we were informed that Eq. (\ref{DCPV}), the
relation between the direct CP asymmetries and CP-averaged branching
ratios of $B^0 \to \pi^+ \pi^-$ and $B^0 \to \pi^- K^+$, has been derived
previously by R. Fleischer \cite{fleischer}.

\end{document}